\documentclass[12pt,a4paper]{article}
\usepackage{blkarray}
\usepackage{amsfonts}
\usepackage{amsmath}
\usepackage{geometry}
\usepackage{graphicx}
\usepackage{amssymb}
\usepackage{abstract}
\usepackage{setspace}
\usepackage{geometry}
\providecommand{\U}[1]{\protect\rule{.1in}{.1in}}

\def\fr{\frac}
\def\be{\begin{equation}}
\def\ee{\end{equation}}
\def\ba{\begin{eqnarray}}
\def\ea{\end{eqnarray}}
\def\pa{\partial}
\def\ra{\rightarrow}
\def\s{\sigma}

\def\e{\varepsilon}
\def\a{\alpha}
\def\b{\beta}
\def\g{\gamma}
\def\d{\delta}
\def\t{\tau}

\def\pt{\phantom{a}}

\def\na{\nabla}

\begin{document}

\title{Principles of \textbf{A}rrangement \textbf{F}ield \textbf{T}heory}

\author{Diego Marin \thanks{dmarin.math@gmail.com}}

\maketitle

\begin{abstract}\begin{spacing}{1.2}
\noindent In this paper I attempt to summarize the fundamental principles which underlie to \emph{Arrangement Field Theory}. In my intention the exposition would be the most possible intelligible and self-contained. However the exposed concepts are revisited in the light of the new researches, so that they could appear slightly different than in the previous works. Much emphasis is posed here to the power of theory to predict the number of fermionic families (flavours) and space-time dimensions. I also give a quick glance to the entanglement phenomenon and its interpretation as microscopic wormhole.
\end{spacing}
\end{abstract}

\newpage
\tableofcontents
\newpage

\section{Introduction}

In the beginning of 2012 I've start the spreading of several ideas for the construction of a new Theory of Everything which could be called \emph{Arrangement Field Theory}. Unfortunately, academic world have ignored all such ideas both because the author is not affiliated to any university and because \emph{Arrangement Field Theory} (from now \emph{AFT}) is in slight contrast with \emph{String Theory} and \emph{Loop Quantum Gravity}.

For example \emph{AFT} is not constructed in a space-time with a preset number of dimensions. Dimensionality here is a free parameter whose most probable value is determined by theory itself. The theory considers in fact space-time as an abstract ensemble of \lq\lq atoms'', intended here as the smallest components (of minimal iper-volume) in which the space-time can be fragmented. In a similar way \emph{AFT} predicts the number of families (flavours) of fermionic fields compatible with a given dimensionality (for $d=4$ it gives $1$ or $3$ families). See section \ref{dimensions}.

The fundamental function of theory defines, for any couple of \lq\lq atoms'', the probability for finding them one beside the other. See section \ref{preliminary}. The shape of universe and the localization of its components assume then a dynamical character, oscillating freely around a \lq\lq middle'' configuration which is the one perceived in daily life. In this framework, the Quantum Entanglement phenomenon between two particles is explained as the annulment of distance between the two particles when this is measured along an extra dimension which doesn't appear in the middle configuration. See section \ref{entanglement}. The phenomenon becomes then the quantum version of wormhole, where every particle assumes characters of a microscopic black hole.

In the continuous limit, \emph{AFT} includes most features of great unification theories based upon gauge group $SU(6)$. See sections \ref{gauge} and \ref{representation}. However it does't throw away \emph{String Theory} at all. In the first of my papers \cite{Quat1}, although pleonastic because of new concepts exposed here (sections \ref{preliminary}, \ref{curves}, \ref{congruences}), it preserves anyway a good section focused on a plausible triality between \emph{AFT, String Theory} and \emph{Loop Quantum Gravity}.

\section{Preliminary definitions}
\label{preliminary}

We start by giving the eight pillars of \textbf{A}rrangement \textbf{F}ield \textbf{T}heory:

\begin{itemize}
\item We define the physical space $\Lambda$ (possibly a space-time) as an abstract ensemble of \lq\lq space atoms'' labeled with Latin letters, i.e. $\Lambda = \{i,j,l,u,v,w,\ldots\}$ with $i,j,l,u,v,w,\ldots$ atoms of space;
\item $\Lambda$ is a topological space with discrete topology;
\item For every couple $i,j \in \Lambda$ we associate an element $M_{ij}$ in some $C^\ast$-algebra $\mathfrak{H}$;
\item An associated graph $\Gamma_\Lambda$ is an oriented abstract graph whose nodes are in one to one with space atoms $i,j,l,u,v,w,\ldots$ and any arrow which goes from node $i$ to node $j$ is labeled by the corresponding $M_{ij}$;
\begin{figure}[h]
\centering\includegraphics[width=0.5\textwidth ]{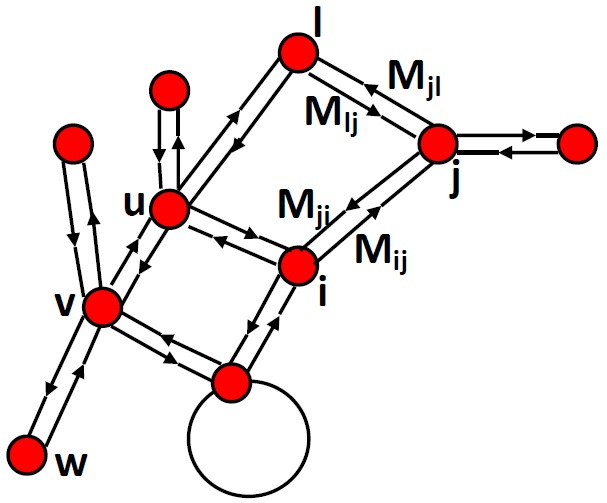}\caption{An example of associated graph}
\label{grafo}
\end{figure}
\item A non drawn arrow between $i$ and $j$ would correspond to $M_{ij}=0$;
\item We define a norm for the associated graph $\Gamma_\Lambda$ as $||\Gamma_\Lambda|| = max_{ij} ||M_{ij}||$;
\item $P_{ij} = \fr{M_{ij}}{||\Gamma_\Lambda||}$ is understood as the probability amplitude for the atom $i$ to be next to (or to be connected with) the atom $j$;
\item Note that atom $i$ can be connected to atom $j$ without $j$ is connected to $i$. This character could be good for describe black holes horizons, where exterior is connected to interior but reverse isn't true.
\end{itemize}

\section{Curves \& covariant derivatives}
\label{curves}

A \textbf{curve} $\g$ in $\Lambda$ is an ordered sequence of atoms. Ex.:

$$\g = \{l,u,v,w\} \qquad with \qquad l<u<v<w.$$

\noindent In this case we can say that $u$ precedes $l$ along $\g$ or $l$ follows $u$ along $\g$. For every curve $\g$ in $\Lambda$ we can define a \textbf{covariant derivative operator} $\na[\g]$ as follows:

$$\left(\na[\g]\right)_{ij} = \left\{\begin{array}{lcc}M_{ij} & if & i,j\in\g,\quad i<j,\quad \nexists l | i<l<j \\
                            0 & otherwise & \end{array}\right.$$

\noindent By defining $\bar{\g}$ as the same of $\g$ with reverse order, we can explicit $\na[\g]+\na[\bar{\g}]$ by using a simple matricial representation:
$$\na[\g]+\na[\bar{\g}] =$$
$$\left(\begin{array}{cccccccc}
\ddots & \ddots & \ddots & \ddots & \ddots & \ddots & \ddots & \ddots \\
\ddots & 0 & e_{-2}^{\g}(A^{-2}_\g + 1) & 0 &0 & 0 & 0 & \ddots \\
\ddots & -e_{-1}^\g & 0 & e_{-1}^\g (A^{-1}_\g +1)& 0 & 0 & 0 & \ddots \\
\ddots & 0 & -e_{0}^\g & 0 & e_{0}^\g (A^{0}_\g +1)& 0 & 0 & \ddots \\
\ddots & 0 & 0 & -e_{1}^\g & 0 & e_{1}^\g (A^{1}_\g +1)& 0 & \ddots \\
\ddots & 0 & 0 & 0 & -e_{2}^\g & 0 & e_{2}^\g (A^{2}_\g +1)& \ddots \\
\ddots & 0 & 0 & 0 & 0& -e_{3}^\g & 0 & \ddots \\
\ddots & \ddots & \ddots & \ddots & \ddots & \ddots & \ddots & \ddots \\
\end{array}\right)$$
\be \pt\pt\pt \label{matrice} \ee

\noindent The enumeration of space-time atoms, from which the arrangement of rows and columns in $\na[\g]+\na[\bar{\g}]$ is derived, is made to coincide here with the order inside $\g$. The introduction of fields $e^\g$ and $A_\g$ is nothing but a free and useful choice for parameterizing $\na[\g]$. For their role in the following we call $e^\g$ \lq\lq vielbein field'' and $A_\g$ \lq\lq gauge-gravitational field''. We'll see that internal indices of $e_\g$ and $A^\g$ (i.e. indices in the tangent space) will appear as internal indices of $\mathfrak{H}$ generators.

We conclude by exhibiting the continuous limit of $\na[\g]+\na[\bar{\g}]$. Let start by consider a field $V(n)$ defined at $n$ regular intervals separated by length $\Delta$ along a straight line or a circle (where the last and the first point of evaluation coincide). Discrete derivative along line or circle can be expressed as
$$ \pa V(n) = \fr {V(n+1)-V(n-1)}{2\Delta} $$
$$ \left(\begin{array}{c}
\ddots \\
\pa V(0) \\
\pa V(1) \\
\pa V(2) \\
\pa V(3) \\
\ddots
\end{array}\right) = \fr 1{2\Delta}\left(\begin{array}{cccccc}
\ddots & \ddots & \ddots & \ddots & \ddots & \ddots\\
\ddots & 0 & 1 & 0 & 0 & \ddots \\
\ddots & -1 & 0 & 1 & 0 & \ddots \\
\ddots & 0 & -1 & 0 & 1 & \ddots\\
\ddots & 0 & 0 & -1 & 0 & \ddots\\
\ddots & \ddots & \ddots & \ddots & \ddots & \ddots
\end{array}\right)\left(\begin{array}{c}
\ddots \\
V(0) \\
V(1) \\
V(2) \\
V(3) \\
\ddots
\end{array}\right)$$

\noindent By using the same convention
$$\na[\g]+\na[\bar{\g}] \longrightarrow 2\Delta e^\g (x) \left(\pa_\g + A_\g(x)\right)$$

\noindent At this level the concept of distance is not defined yet. Hence $\Delta$ has to be here a fundamental length to be measured through experiments. You'll see in section \ref{dimensions} that $\Delta$ has to be close to the Planck length, i.e.

$$\Delta \approx 1,6 \cdot 10^{-35} \quad meters$$

\section{Congruences \& dimensionality}
\label{congruences}

\noindent A family of curves $\theta = \{\g^1,\g^2,\ldots,\g^n\}$ is a \textbf{congruence} in $\Lambda$ if
\ba && \g^a \cap \g^b = 0 \qquad \forall a,b \quad with \quad a \neq b \nonumber \\
    && \bigcup_a \g^a = \Lambda \nonumber \ea

\noindent Two congruences $\theta_1$ and $\theta_2$ are \textbf{independent} if
\ba && \left(\na[\g^a]\right)_{ij} \neq 0 \Rightarrow \left(\na[\g^b]\right)_{ij} = 0 \nonumber \\
    && \left(\na[\g^b]\right)_{ij} \neq 0 \Rightarrow \left(\na[\g^a]\right)_{ij} = 0 \nonumber \\
    && \qquad\qquad\qquad\qquad for\pt any \pt \g^a \in \theta_1, \quad \g^b \in \theta_2, \quad i,j \in \Lambda \nonumber \ea

\noindent Remember that in the matrix (\ref{matrice}) we have ordered rows and columns according to the order in $\g$. This choice can't be accomplished simultaneously for two or more curves belonging to independent congruences, because they have necessarily two different orders. So any sum of type $\na[\g^a] + \na[\g^b]$, with $\g^a,\g^b$ belonging to independent congruences, will return a matricial representation with a more complex structure than (\ref{matrice}).

Definition of $\na[\_]$ can be trivially extended to congruences $\theta$:
$$\left(\na[\theta]\right)_{ij} = \sum_{\g \in \theta} \left(\na[\g]\right)_{ij} $$

\noindent The idea of congruence sends straightforward to the concept of \textbf{Dimensionality}. In fact, \textbf{Dimensionality} of $\Lambda$ is the minimal number $n$ of independent congruences $\theta_1,\theta_2,\ldots,\theta_n$ for which the following relation is satisfied:
$$\sum_{a=1}^n \sum_b \left(\na[\g_a^b]\right)_{ij} + \left(\na[\bar{\g}_a^b]\right)_{ij} = M_{ij} \qquad \forall i,j \in \Lambda \quad with \quad i\neq j$$

\noindent Here index $a$ runs over congruences, while $b$ runs over curves inside a single congruence. In the continuous limit
$$\sum_{indep.\theta}\left(\na[\theta]+\na[\bar{\theta}]\right) \longrightarrow 2\Delta \sum_{indep.\theta} e^{\theta} (x) \left(\pa_{\theta} + A_{\theta}(x)\right)$$

\noindent or simply
$$\sum_{indep.\theta}\na[\theta]+\na[\bar{\theta}] \longrightarrow 2\Delta e^{\theta} (x) \left(\pa_{\theta} + A_{\theta}(x)\right)$$

\noindent if we use Einstein convention on repeated indices. Here we have used
$$e^{\theta} = \sum_{\g \in \theta} e^{\g} \qquad\qquad A_{\theta} = \sum_{\g\in \theta} A_{\g}$$

\section{Entanglement}
\label{entanglement}

We can expand any polynomial $P^{n}(M)$ of degree $n$ inside action around a middle configuration with $d$ dimensions:
$$tr\,P^n\left(M\right) = tr\,P^n\left(\sum_{indep.\theta}\left(\na[\theta]+\na[\bar{\theta}]\right)+\phi\right)\approx$$
$$\approx tr\,P^n\left(\langle M \rangle\right)+\sum_{i,j,k}P^{n-m-1}\left(\langle M \rangle\right)^{ij}
E^{jk}P^{m}\left(\langle M \rangle\right)^{ki}+O\left(E^2\right) $$

\noindent where
$$E = M - \langle M \rangle = \sum_{\theta \neq \theta_1,\theta_2,\ldots,\theta_d}\left(\na[\theta]+\na[\bar{\theta}]\right) -\phi$$

\noindent and $\phi$ is here the diagonal piece of $M$, i.e. $\phi^{ij} = \d^{ij}M^{jj}$. By substituting
$$tr\,P^n\left(M\right) \approx tr\,P^n\left(\sum_{b=1,\ldots,d}\left(\na[\theta_b]+\na[\bar{\theta}_b]\right)+\phi\right)+$$
$$\sum_{i,j,k}P^{n-m-1}\!\left(\sum_{c=1,\ldots,d}\!\!\left(\na[\theta_c]+\na[\bar{\theta}_c]\right)+\phi\right)^{ij}
\!\!\!\!E^{jk}P^{m}\!\left(\sum_{f=1,\ldots,d}\!\!\left(\na[\theta_f]+\na[\bar{\theta}_f]\right)+\phi\right)^{ki}
\!\!\!\!+\ldots $$

\noindent In the continuous limit
$$\fr 1{(2\Delta)^n} tr\,P^n\left(M\right) \longrightarrow \int dx\,P^n\left(e^{b} (x) \left(\pa_{b} + A_b(x)\right)+\phi(x)\right)+$$
$$\int\!\! dx dy P^{n-m-1}\!\left(e^{c} (x) \left(\pa_{c} + A_c(x)\right)+\phi(x)\right)
E(x,y)P^{m}\!\left(e^{f} (y) \left(\pa_{f} + A_f(y)\right)+\phi(y)\right)+\ldots $$

\noindent To include quantum perturbations we have thus to consider a non local field $E(x,y)$ or $E^{ij}$. This is because two atoms $i,j$ can be located far away in the medium (classical) configuration, while appearing as neighbors in some other configurations. We call the non local field $E(x,y)$ \textbf{entanglement field} inasmuch it appears to be useful to describe (non local) entanglement phenomena.

Note that $\na[\g]$ describes connections between neighboring atoms (better it determines what atoms are neighbors and what not), but at the same time it gives a linear momentum $P_\g$ along $\g$. Conversely $\phi^{ii}$ describes a connection from an atom to itself. Moreover it can be chosen in such a way that it takes values in the $su(2)$ algebra (by choosing $\mathfrak{H} \supset su(2)$). Hence it can describe a spin operator. This gives a completely new understanding of spin as a linear momentum along \lq\lq pointwise loops''.

However in what follows we'll concentrate on the local piece of action neglecting the influence of entanglement except as regards the states counting.

\section{The algebra $\mathfrak{H}$}

We choose $\mathfrak{H} = gl(5, \mathbb{C})$, i.e. the group of linear transformations inside $\mathbb{C}^5$ so parameterized
$$gl(5, \mathbb{C}) \ni \Omega = \left(\begin{array}{c|c}
  sl(2,\mathbb{C}) & \begin{array}{ccc} (\psi_R^1)^c &  (\psi_R^2)^c & (\psi_R^3)^c \end{array} \\
  \hline
  \begin{array}{c} \psi_L^1 \\  \psi_L^2 \\  \psi_L^3 \end{array} & sl(3,\mathbb{C}) \end{array}\right)+\left(\a+i\b\right)\mathbf{1}_5+$$
$$+(\g+i\d)\left(\begin{array}{ccccc}
  3i & 0 & 0 & 0 & 0 \\
  0 & 3i & 0 & 0 & 0 \\
  0 & 0 & -2i & 0 & 0 \\
  0 & 0 & 0 & -2i & 0 \\
  0 & 0 & 0 & 0 & -2i
  \end{array}\right)$$

\noindent $\a,\b,\g,\d \in \mathbb{R}$. $^c$ denotes charge conjugation. $gl(5, \mathbb{C})$ can be consider as a ring of iper-complex numbers with a real unit $\mathbf{1}_5$ (also called $T^0$) and $49$ imaginary units ($T^1,T^2,\ldots,T^{49}$). Every product between such units is deduced from matricial product. Note that $gl(5, \mathbb{C})$ is closed with respect to such product, while other algebras (like for example $sl(5, \mathbb{C})$) are not closed with respect to the same.

In this way $A_\theta$ contains a component in $sl(2,\mathbb{C})$ which is suitable to describe gravity. Also $\phi$ contains components in representation $\mathbf{2}$ respect $sl(2,\mathbb{C})$ and $\mathbf{3}$ respect $sl(3,\mathbb{C}) \sim su_L(3) \oplus su_R(3)$, so that it gives account for spinors left and right in three families.

For what follows is useful to calculate the number of bosonic generators ($b = 2*(3^2-1)+2*(2^2-1) + 4 = 26$) and fermionic generators ($f = 2*2*(2*3) = 24$). Much important for the calculation of space-time dimensions in the middle configuration will be the difference $b-f = 2$.

\section{Gauge fields}\label{gauge}

We suppose that all atoms in $\Lambda$ are superimposed in groups $W$ of $m$ elements:
\ba && W^a = \{i^a_1,i^a_2,i^a_3,\ldots,i^a_m\} \nonumber \\
    && \bigcup_a W^a = \Lambda \nonumber \ea

\noindent In this case we define a curve $\g$ as an ordered sequence of superimpositions. Ex.:
$$\g = \{W^2,W^{15},W^{24},W^{127}\} \qquad\qquad W^2 < W^{15} < W^{24} < W^{127}$$

\noindent In this condition every element in the matrix (\ref{matrice}) will be replaced by a matrix $m \times m$. If we consider the effects of a local $GL(5,\mathbb{C})$ symmetry inside a single $W^a$, we can interpret it as an overall $GL(5,\mathbb{C})^m$ symmetry. However, if the physics doesn't depend from the structuring of points $i^a_b$ inside $W^a$, this symmetry expands to $GL(5m,\mathbb{C})$, more or less as happens in \emph{String Theory} with the superimposition of $m$ D-branes. Finally, if we consider the groups $W$ as the real physical points (or events) we can speak about a local $GL(5m,\mathbb{C})$, clearly referring \lq\lq local'' to groups and not to the single atoms.

In this way, for every group $W^a$ we can write $A_\theta (i^a,j^a)$ as $A_\theta^{ij}(a)$ and $e^\theta (i^a,j^a)$ as $\d_{ij}e^\theta(a)$. $A_\theta(a)$ is then intended as an element in the algebra $gl(5m, \mathbb{C}) = gl(m, \mathbb{C}) \otimes gl(5, \mathbb{C})$. Note that is $\otimes$ at algebra level (and not $\oplus$) so that for the group $GL(5m, \mathbb{C})$ the same factorization with $\otimes$ doesn't work. Finally note that the $2\times 2$ submatrix of $e^\theta(a)$ can be written as $\sum_{\t=0}^3 e^\theta_{\t}(a)\s^\t$ with $e^\theta_{\t} \in \mathbb{C}$. Hence for $dim.\Lambda =4$ we have that $e^\theta_{\t}$ is a tetrad field, although not for Minkowski space but conversely for its complexification.

It's reasonable that superimposed points has to have an unique tangent space, thus reducing the algebra $gl(m, \mathbb{C}) \otimes gl(5, \mathbb{C})$ to $gl(m, \mathbb{C}) \oplus gl(5, \mathbb{C})$. Only the now reduced group factorizes in $GL(m, \mathbb{C}) \otimes GL(5, \mathbb{C})$. Moreover, every transformation $\Omega \in gl(m, \mathbb{C})$ which acts inside a single $W^a$ has to not change the norm of $A$, i.e. $A^i_{\theta i} = (e^\Omega A e^{\Omega^\dag})^i_{\pt i}$ so that $\Omega$ has to belong to $u(m)=u(1)\oplus su(m)$. In the following we impose $m=6$ by suggesting a chain of symmetry breakages:

\begin{itemize}
\item $gl(30,\mathbb{C})\curvearrowright$
\item $gl(6,\mathbb{C})\oplus gl(5,\mathbb{C})\curvearrowright$
\item $(u(1)\oplus su(6))\oplus (sl(2,\mathbb{C})\oplus sl(3,\mathbb{C})\oplus (\mathbf{2}\otimes \mathbf{3})
\oplus (\bar{\mathbf{2}} \otimes \bar{\mathbf{3}})\oplus u(1) \oplus u(1))\curvearrowright$
\item $su(3) \oplus su(2) \oplus 4 u(1) \oplus sl(2,\mathbb{C})\oplus sl(3,\mathbb{C}) + (\mathbf{2}\otimes \mathbf{3})\oplus (\bar{\mathbf{2}} \otimes \bar{\mathbf{3}})$
\end{itemize}

\noindent So we have included all the standard model: $su(3)$ (cromodynamic), $su(2)\oplus u(1)$ (electroweak), $sl(2,\mathbb{C})$ (gravity), $sl(3,\mathbb{C}) \sim su_L(3)\oplus su_R(3)$ (flavour), $\mathbf{2} \otimes \mathbf{3}$ (fermions in three flavours) and $\bar{\mathbf{2}} \otimes \bar{\mathbf{3}}$ (anti-fermions in three flavours).

\section{The supporting space $\Lambda'$}

\begin{figure}[h!]
\centering\includegraphics[width=0.2\textwidth ]{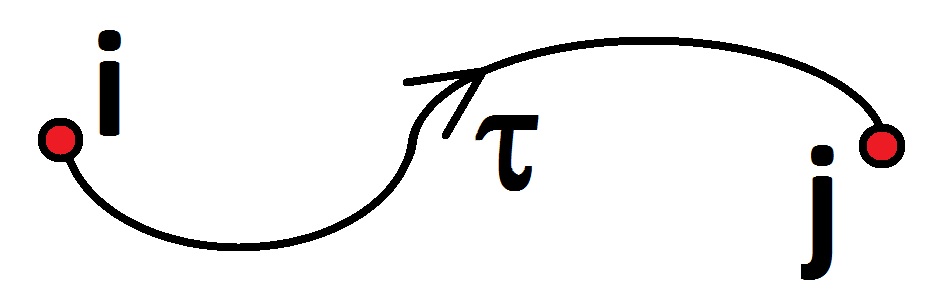}
\caption{Parametrization of the arrow which goes from atom $i$ to atom $j$.}\label{string}
\end{figure}

Let's start to parametrize any arrow $i\ra j$ in the graph $\Gamma_\Lambda$ through a parameter $\t_{ij}$ which runs from $0$ to $1$ like in figure $\ref{string}$.

\noindent Then we define a current $j^{ij}(\t_{ij})$ with values in $gl(5,\mathbb{C})$ associated with the path $i\ra j$. Such current has to transform in the covariant way under reparametrizations $\t'(\t)$ with $\t'(0) = 0$ and $\t'(1) = 1$:
$$j^{\prime ij}(\t'_{ij}) = j^{ij}(\t_{ij})\left(\fr{d\t'_{ij}(\t_{ij})}{d\t_{ij}}\right)^{-1}$$

\noindent We see that the product $j^{ij}(\t_{ij}) d\t_{ij}$ doesn't depend from the chosen parametrization, so that we can define any element $M^{ij}$ as the integration of $j^{ij}$ along the arrow $i\ra j$:
$$M^{ij} = \int_0^1 \fr{d\t_{ij}}{2\pi} j^{ij}(\t_{ij})$$

\noindent Thanks to parametrization invariance there is no need to consider the arrow as a physical real existing object.

In the previous section we have seen that the vielbein field $e_\theta$ is a complex entity, a fact which suggests the existence of many imaginary dimensions in a number which equals the number of real dimensions. We suppose that any imaginary dimension is closed in such a way to draw a microscopic circle parametrized by a coordinate $y$ which runs from $0$ to $1\approx 0$. Any imaginary dimension can be used to define the diagonal elements of $M$ like $M^{ii}$:
$$M^{ii} = \oint_0^{0\approx 1} \fr{dy_{i}}{2\pi} j^{i}(y_{i})$$

\noindent Let's define the exponentiated version $\mathfrak{M}^{ij}$ of $M^{ij}$ where:
$$\mathfrak{M}^{ij} = exp\,M^{ij} = exp\,\int_0^1 \fr{d\t_{ij}}{2\pi} j^{ij}(\t_{ij})$$

\noindent Consider now a closed curve $\g =\{i,j,k,l,\ldots,i\}$ like the one in figure \ref{surface}:

\begin{figure}[h!]
\centering\includegraphics[width=0.3\textwidth ]{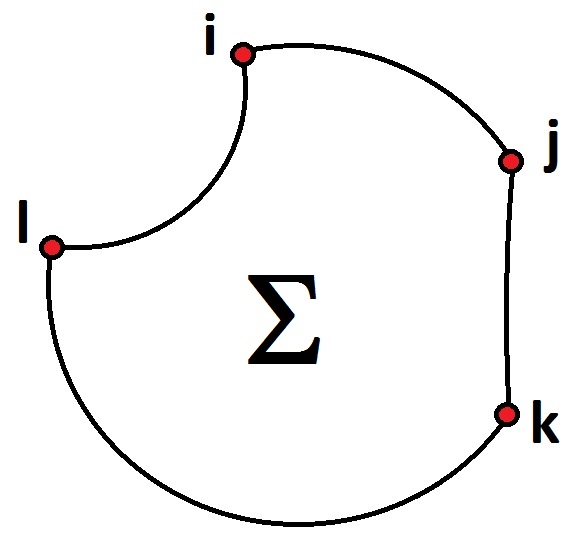}
\caption{A closed curve which touches four atoms}\label{surface}
\end{figure}

\noindent Consider then the action
$$S = \mathfrak{M}^{ij} \mathfrak{M}^{jk} \mathfrak{M}^{kl} \ldots \mathfrak{M}^{\ldots i}
=exp\,\oint_\g \fr{d\t}{2\pi} \tilde{j}(\t)$$
where
$$\tilde{j} = \begin{array}{ll} j^{ij} & \text{from $i$ to $j$} \\ j^{jk} & \text{from $j$ to $k$}\\ j^{kl} & \text{from $k$ to $l$} \\ \ldots & \ldots \\ j^{\ldots i} &\text{from $\ldots $ to $i$}\end{array}$$

\noindent We can introduce whatever surface $\Sigma$ which possesses $\g$ as boundary; we can choose to parameterize $\Sigma$ by means of a couple of coordinates $\s_1,\s_2$ or alternatively by a complex coordinate $z=\s_1+i\s_2$. Moreover we define the value of $\tilde{j}(z)$ in any internal point of $\Sigma$ by using the Cauchy formula
$$\tilde{j}(z)= \sum_a \tilde{j}^a(z) T^a \quad\qquad \begin{array}{c} \tilde{j}^a \in \mathbb{C}\\ T^a \in gl(5,\mathbb{C})\end{array} \quad\qquad \tilde{j}^a(z) = \oint_\g \fr{dz'}{2\pi i}\fr{\tilde{j}^a (z')}{z'-z}$$

\noindent which is the unique way to have $\tilde{j}^a$ holomorphic in $\Sigma$. At this point the action can be elaborated by using the Gauss theorem
\ba S &=& exp\,\oint_\g \fr{d\t}{2\pi} \tilde{j}(\t)
= exp\,\fr {1}{2\pi i}\int_\Sigma dz d\bar{z} \, \pa_z \tilde{j}(z)\nonumber\\
&=& exp\,\fr {1}{2\pi i}\int_\Sigma d^2\s \left\{\left[\pa_1 \tilde{j}(z)_1 + \pa_2 \tilde{j}(z)_2\right]+i\left[\pa_1 \tilde{j}(z)_2 - \pa_2 \tilde{j}(z)_1\right]\right\} \nonumber\\
&=& exp\,\fr {1}{2\pi i}\int_\Sigma d^2\s \left\{\d^{ab}g_{ab}+i\e^{ab}f_{ab}\right\} \nonumber\ea

\noindent where $\tilde{j}(z)_1 = Re\, \tilde{j}(z)$, $\tilde{j}(z)_2 = Im\,\tilde{j}(z)$, $g_{ab} = \fr 12 \left[\pa_a\tilde{j}(z)_b + \pa_b\tilde{j}(z)_a\right]$ and $f_{ab} = \fr 12 \left[\pa_a\tilde{j}(z)_b - \pa_b\tilde{j}(z)_a\right]$. The same process can be accomplished for surfaces $\Sigma$ which have the over mentioned circles (the ones in the imaginary directions) as boundary.

Note that the action $S$ does't depend from the chosen $\Sigma$ because a change in $\Sigma$ is exactly compensated by the change of $j(z)$ induced by Cauchy formula. This is the well known \lq\lq Conformal Invariance''. For the same motive there is no need to consider the surface $\Sigma$ as a physical real existing object.

The definition of \lq\lq supporting space'' naturally comes $\Longrightarrow$\\
$\Longrightarrow$ A space $\Lambda'$ is a \textbf{supporting space} for the graph $\Gamma_\Lambda$ when the following requirements are satisfied:
\begin{itemize}
\item There exists a family of surfaces $\Sigma_i \subset \Lambda'$ with $\bigcup_i \pa \Sigma_i = \Gamma_\Lambda$ and $\pa \Sigma_i \nsubseteq \pa \Sigma_j$ for every couple $i,j$;
\item If another space $\Lambda''$ satisfies the above requirement, then $dim\,\Lambda'' \leq dim\,\Lambda'$.
\end{itemize}

\noindent It is useful to parameterize $\Lambda'$ by using complex coordinates $\xi^\mu$ ($\mu = 1,2,\ldots dim\,\Lambda'$) appropriately chosen for having restrictions $\xi^\mu(z)|_\Sigma$ holomorphic in every $\Sigma \subset \Lambda'$. This permits to rewrite action as
\ba S &=& exp\,\fr {1}{2\pi i}\int_\Sigma d^2\s \left\{\d^{ab}G_{\mu\nu}\xi^{\mu}_{,a}\xi^{\nu}_{,b}+i\e^{ab}F_{\mu\nu}\xi^{\mu}_{,a}\xi^{\nu}_{,b}\right\} \nonumber\ea

\noindent Here $G_{\mu\nu}$ and $F_{\mu\nu}$ are any metric and antisymmetric background which induce the correct $g_{ab}$ and $f_{ab}$ when restricted to $\Sigma \subset \Lambda'$. In such a way we recover \emph{String Theory} with one fundamental difference: here we have a tree made of strings docked to one another at their ends (a sort of strings-spin-network). A single string doesn't move, but conversely we have the information which moves from a string to another.

It is well known that conformal invariance can be preserved at quantum level only if \emph{String Theory} is constructed in a space with $26$ real dimensions (i.e. $13$ complex dimensions). Hence $26$ is the number of dimensions for the supporting space $\Lambda'$, but $\Lambda'$ is \textbf{not} the physical space (or space-time); specially its number of dimensions has nothing to do with the number of dimensions of the physical space (or space-time). This is in fact the \textbf{BIG ERROR} of \emph{String Theory}: to consider $\Lambda'$ as a real existing physical space (or space-time). $\Lambda'$ has't to be compactified here (except for the cylindrical compactification in the imaginary directions), so that we have removed at all the huge ambiguity which harms \emph{String Theory}. 

Conversely, the dimensionality of $\Lambda'$ poses some limits to the number of nonzero elements in the arrangement-matrix $M$.

\section{How the fermions born}

Consider the definition of diagonal components $M^{ii}$ as the integration of a current $j^i$ along a circular path $C_i$ (in a closed imaginary direction) parameterized by $y_i$:
$$M^{ii} = \oint_{C_i}\fr {dy_i}{2\pi} j^i (y_i).$$

\noindent Express also the non diagonal components $M^{kl}$ with $k\neq l$ as the integration of a current $j^{kl}$ along an open path $\g_{kl}$ between $k$ and $l$ parameterized by $\t_{kl}$:
$$M^{kl} = \int_{\g_{kl}} \fr {d\t}{2\pi} j^{kl} (\t_{kl}).$$

\noindent We require that the supporting space $\Lambda'$ is an orbifold parameterized by complex coordinates $\xi^\mu$ ($\mu=1,2,\ldots,26$) which satisfy the specular identification $\xi^\mu \approx -\xi^\mu$. This means that any holomorphic function $f$ having domain in a Riemann surface $\Sigma \subset \Lambda'$ has to satisfy $f(z) = f(Re\, \xi^\mu(z),Im\,\xi^\mu(z)) = f(- Re\, \xi^\mu(z),- Im\, \xi^\mu(z))$.

\begin{figure}[h!]
\centering\includegraphics[width=0.7\textwidth ]{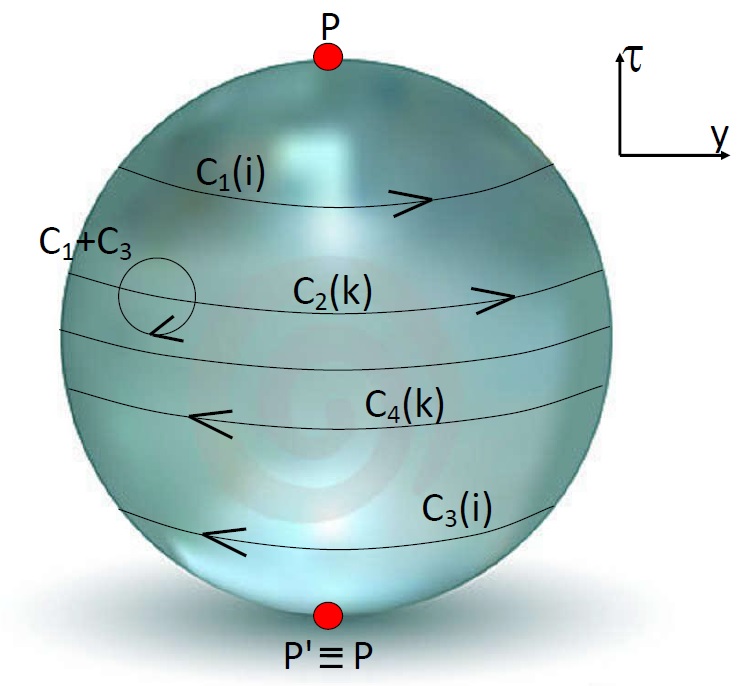}\caption{Orbifold structure of space time $\Lambda'$.}
\label{sfera}
\end{figure}

\noindent For every couple of components $M^{ii}\!\!-\!\!M^{kk}$ (or $M^{ij}\!\!-\!\!M^{kl}$), we can find a Riemann surface $\Sigma \subset \Lambda'$ isomorphic to the one depicted in figure \ref{sfera} which contains the paths of both components. At this point we can determine the commutation relations for the quantum operators corresponding to $M^{ii}$ and $M^{kl}$ (we call them $\hat{M}^{ii}$ and $\hat{M}^{kl}$). Clearly we have
$$\hat{M}^{ij}\hat{M}^{kl} = \hat{M}^{kl}\hat{M}^{ij}$$

\noindent because there is no way for deforming a path from $i$ to $j$ in a path from $k$ to $l$ through a continuous series of infinitesimal transformations. Thus the operators $\hat{M}^{ij}$ describe bosonic degrees. Conversely:
\ba \langle M^{ii}\{C_1\} M^{kk}\{C_2\} + M^{ii}\{C_3\} M^{kk}\{C_2\}\rangle &=& [\hat{M}^{ii},\hat{M}^{kk}]\nonumber \\
&=& \fr 1{2\pi}\fr 1{2\pi i} \oint_{C_2} \oint_{C_1+C_3} dy dz\, \langle j^i(z) j^k(y) \rangle \nonumber \\
&=& \oint_{C_2}\fr{dy}{2\pi} Res_{\e\ra 0}\, \langle j^i(y+i\e)j^k(y)\rangle\neq 0\nonumber \ea

\noindent Note the sign \lq\lq $+$'' instead of \lq\lq $-$'' because paths $C_1$ and $C_3$ have different orientations before identify $P$ with $P'$. This argument works independently from the chosen theory, provided we take $j(z)$ holomorphic in $\Sigma$.

The only role of path-integral (implied in $\langle\pt\rangle $) is to reflect to operators $\hat{M}$ the structure of $\Sigma$ (which says that $C_2$ is comprised between $C_1$ and $C_3$) in such a way to compose a commutator. Moreover
$$ \langle M^{ii}\{C_1\} M^{kk}\{C_2\} - M^{ii}\{C_3\} M^{kk}\{C_2\} \rangle = \{\hat{M}^{ii},\hat{M}^{kk}\}$$

\noindent but $M^{ii}\{C_1\} = M^{ii}\{C_3\}$, so that
$$\{\hat{M}^{ii},\hat{M}^{kk}\} = \langle M^{ii}\{C_1\} M^{kk}\{C_2\} - M^{ii}\{C_1\} M^{kk}\{C_2\} \rangle = 0$$

\noindent Thus the diagonal fields $M^{ii}$ anti-commute and so they describe fermionic degrees.

At this point we see that also the field $\psi_\theta$ describes bosonic degrees, although it transforms in the representation $\left(\fr 12 \oplus \fr 32\right)$ under the action of Lorentz group. Thus it is a ghost field and it has to be removed from the physical states.

Similarly, any scalar field $\phi$ describes fermionic degrees, although it transforms in the representation $0$ under the action of Lorentz group. Thus it is a ghost field too and it has to be removed from the physical states. For this reason, the Higgs boson, if exists, can't has spin $0$.\\

\noindent \textbf{Take a look to the following images. We can trace a correspondence between quantum fields and conic sections. In this way it becomes glaring that fermionic and bosonic fields are different sections of a single entity: the cone. You see that the supporting space $\Lambda'$ contains a privileged point in $\xi^\mu = 0$ which is the vertex of all cones. However every measurable quantity has an expectation value different from zero only if it contains an even product of fermionic fields. Thus every measurable quantity \lq\lq lives'' in the Riemann Sphere, where there aren't special points.}

\begin{figure}[h!]
\centering\includegraphics[width=0.7\textwidth ]{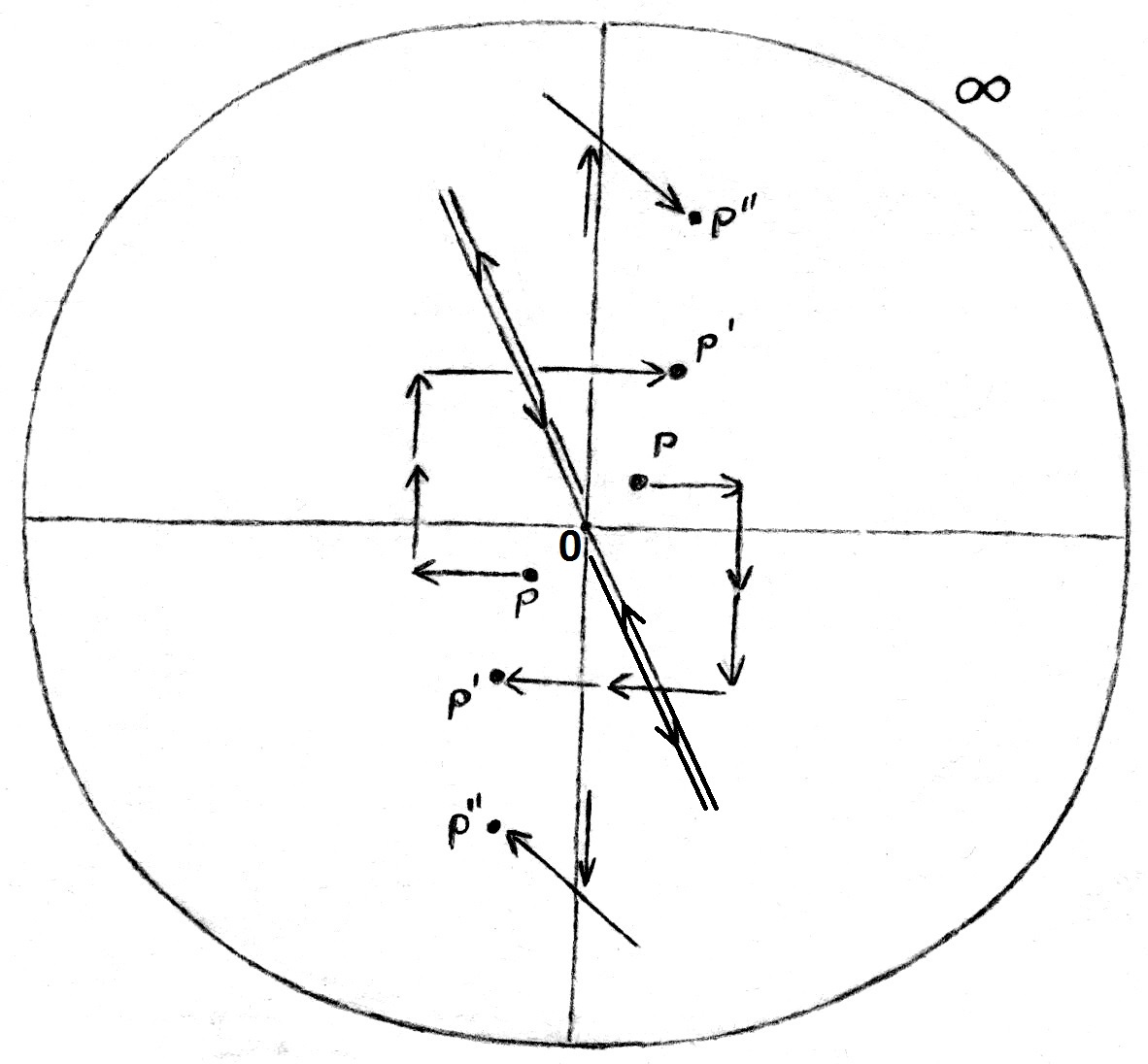}\caption{Points and trajectories in the complex plane (Riemann Sphere) under the identification $z \sim -z$}
\label{S-Riemann}
\end{figure}
\begin{figure}[h!]
\centering\includegraphics[width=0.4\textwidth ]{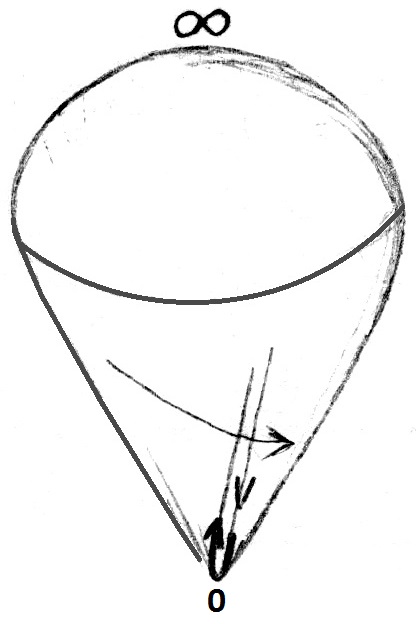}\caption{The Riemann Sphere after the identification $z\sim -z$ becomes omeomorphic to a semicone with only one exception: the point $z= 0$. In fact, geodetics which pass through $z=0$ in the Sphere, become singular trajectories in the semicone which bounce in the singularity $z=0$.}
\label{S-Cone}
\end{figure}
\begin{figure}[h!]
\centering\includegraphics[width=\textwidth ]{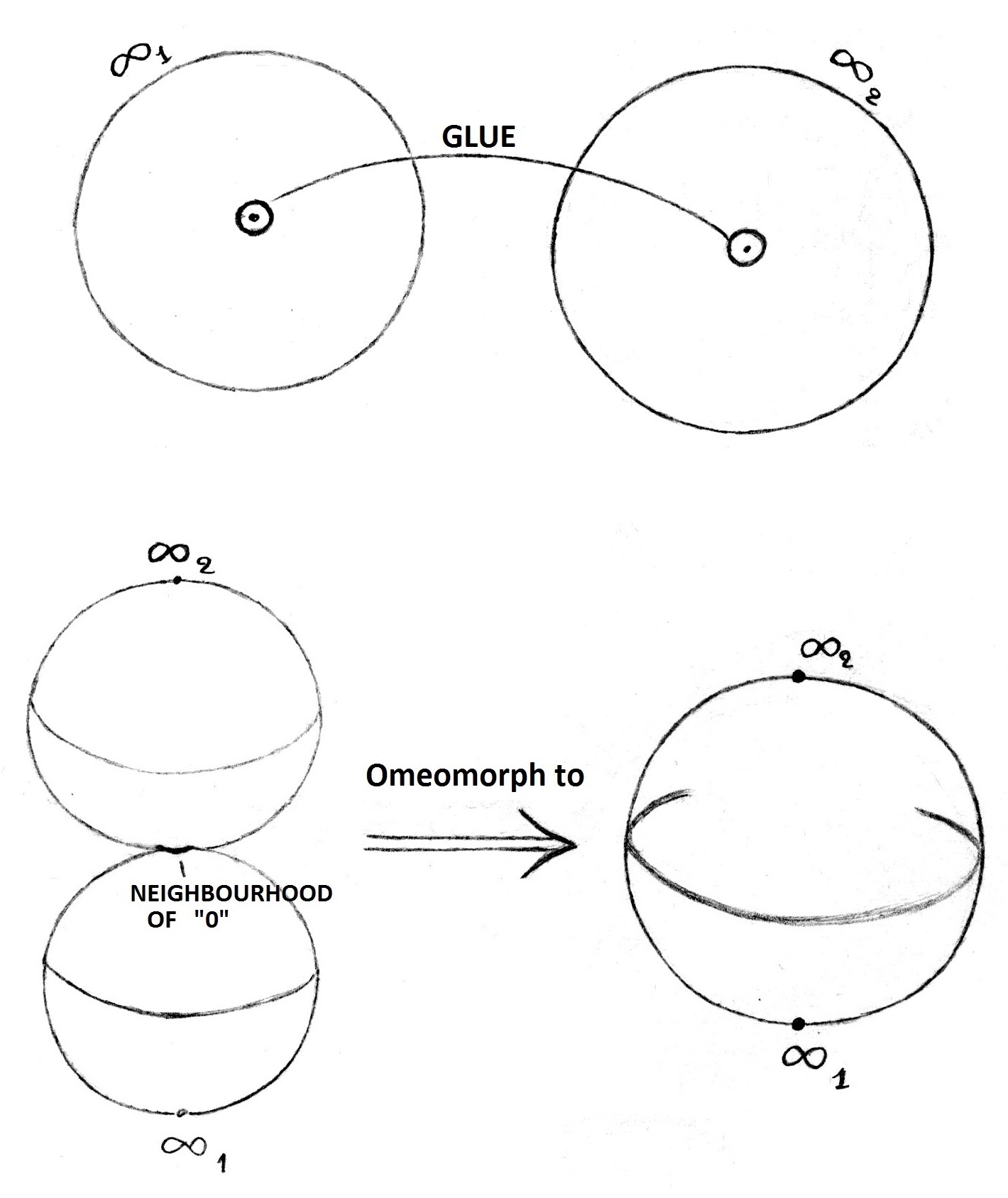}\caption{We can remove the singolarity by gluing together two Riemann spheres deprived of $z = 0$ in the neighbourhood of such point.}
\label{Glue}
\end{figure}
\begin{figure}[h!]
\centering\includegraphics[width=0.3\textwidth ]{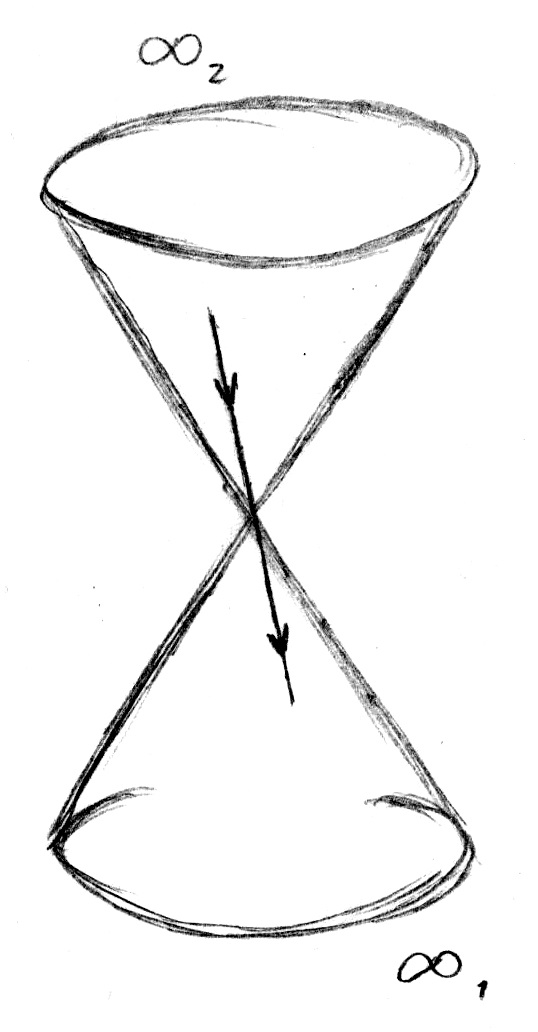}\caption{The \lq\lq double'' Riemann Sphere after the identification $z\sim -z$ becomes omeomorphic to the cone. In this case we have a complete correspondence between geodetics in the sphere and geodetics in the cone.}
\label{Cone}
\end{figure}
\begin{figure}[h!]
\centering\includegraphics[width=0.7\textwidth ]{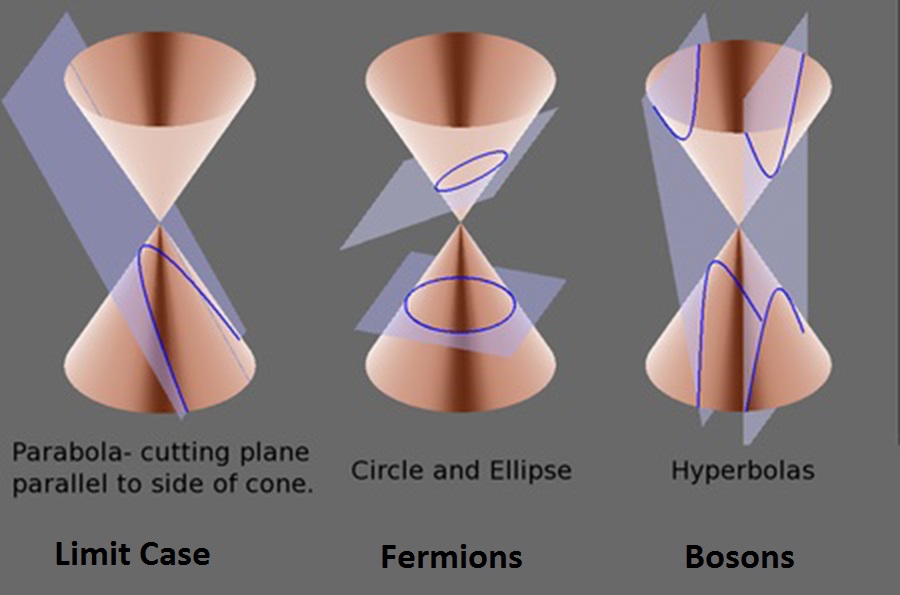}\caption{Correspondence between quantum fields and conic sections.}
\label{conic-sections}
\end{figure}

\newpage
\pt
\newpage
\pt
\newpage
\pt
\newpage
\section[Skew-symmetric representation]{The problem of spinor fields which transform in the skew-symmetric representation}\label{representation}

When we deal with grand unification theories or supersymmetric theories, one of the greater problems we encountered is that gauge fields transform in the adjoint representation of gauge group, while fermions transform in the skew symmetric. In this section we'll see that this problem is only a false problem which spontaneously slips away.

Start by nothing that, when we construct the gauge group $su(6)$, we have at least two choices for the imaginary unit. The first is obviously $i\mathbf{1}_5$, but a second exists and is

$$I = \left(\begin{array}{ccccc}
i & 0 & 0 & 0 & 0\\
0 & i & 0 & 0 & 0\\
0 & 0 & -i & 0 & 0\\
0 & 0 & 0 & -i & 0\\
0 & 0 & 0 & 0 & -i
\end{array}\right)$$
which commutes both with $SL(2,\mathbb{C})$ and $SL(3,\mathbb{C})$ generators. Moreover $I^\dag = -I$ and $I^2 = -1$ so that he has all the features of an imaginary unit. Since $SL(5,\mathbb{C})$ is broken in $SL(2,\mathbb{C})\otimes SL(3,\mathbb{C})$, we can construct the gauge group $su(6)$ by using this unit. Consider now that every spinorial component is expressed through a complex combination of the following generators:

$$\left(\begin{array}{ccccc}
0 & 0 & 0 & 0 & 0\\
0 & 0 & 0 & 0 & 0\\
1 & 0 & 0 & 0 & 0\\
0 & 0 & 0 & 0 & 0\\
0 & 0 & 0 & 0 & 0
\end{array}\right),
\left(\begin{array}{ccccc}
0 & 0 & 0 & 0 & 0\\
0 & 0 & 0 & 0 & 0\\
0 & 0 & 0 & 0 & 0\\
1 & 0 & 0 & 0 & 0\\
0 & 0 & 0 & 0 & 0
\end{array}\right),
\left(\begin{array}{ccccc}
0 & 0 & 0 & 0 & 0\\
0 & 0 & 0 & 0 & 0\\
0 & 0 & 0 & 0 & 0\\
0 & 0 & 0 & 0 & 0\\
1 & 0 & 0 & 0 & 0
\end{array}\right),$$
$$\left(\begin{array}{ccccc}
0 & 0 & 0 & 0 & 0\\
0 & 0 & 0 & 0 & 0\\
0 & 1 & 0 & 0 & 0\\
0 & 0 & 0 & 0 & 0\\
0 & 0 & 0 & 0 & 0
\end{array}\right),
\left(\begin{array}{ccccc}
0 & 0 & 0 & 0 & 0\\
0 & 0 & 0 & 0 & 0\\
0 & 0 & 0 & 0 & 0\\
0 & 1 & 0 & 0 & 0\\
0 & 0 & 0 & 0 & 0
\end{array}\right),
\left(\begin{array}{ccccc}
0 & 0 & 0 & 0 & 0\\
0 & 0 & 0 & 0 & 0\\
0 & 0 & 0 & 0 & 0\\
0 & 0 & 0 & 0 & 0\\
0 & 1 & 0 & 0 & 0
\end{array}\right),$$
$$\left(\begin{array}{ccccc}
0 & 0 & 1 & 0 & 0\\
0 & 0 & 0 & 0 & 0\\
0 & 0 & 0 & 0 & 0\\
0 & 0 & 0 & 0 & 0\\
0 & 0 & 0 & 0 & 0
\end{array}\right),
\left(\begin{array}{ccccc}
0 & 0 & 0 & 1 & 0\\
0 & 0 & 0 & 0 & 0\\
0 & 0 & 0 & 0 & 0\\
0 & 0 & 0 & 0 & 0\\
0 & 0 & 0 & 0 & 0
\end{array}\right),
\left(\begin{array}{ccccc}
0 & 0 & 0 & 0 & 1\\
0 & 0 & 0 & 0 & 0\\
0 & 0 & 0 & 0 & 0\\
0 & 0 & 0 & 0 & 0\\
0 & 0 & 0 & 0 & 0
\end{array}\right),$$
$$\left(\begin{array}{ccccc}
0 & 0 & 0 & 0 & 0\\
0 & 0 & 1 & 0 & 0\\
0 & 0 & 0 & 0 & 0\\
0 & 0 & 0 & 0 & 0\\
0 & 0 & 0 & 0 & 0
\end{array}\right),
\left(\begin{array}{ccccc}
0 & 0 & 0 & 0 & 0\\
0 & 0 & 0 & 1 & 0\\
0 & 0 & 0 & 0 & 0\\
0 & 0 & 0 & 0 & 0\\
0 & 0 & 0 & 0 & 0
\end{array}\right),
\left(\begin{array}{ccccc}
0 & 0 & 0 & 0 & 0\\
0 & 0 & 0 & 0 & 1\\
0 & 0 & 0 & 0 & 0\\
0 & 0 & 0 & 0 & 0\\
0 & 0 & 0 & 0 & 0
\end{array}\right)$$

\noindent All these anticommute with $I$, so that, indicating any of them with $T$, we have, by considering the action of a gauge transformatio $U \in SU(6)$

$$\psi T \ra U\psi T U^\dag = U \psi (U^\dag)^\ast T = U \psi U^T T$$
so that
$$\psi \ra U \psi U^T.$$

\noindent We conclude that, if gauge fields transform in the adjoint representation $A \ra UAU^\dag$, then fermionic fields transform in the skew symmetric representation. To conclude the section we show the disposition of standard model fermions inside a skew symmetric $6\times 6$ matrix:
$$\psi = \left( \begin{array}[c]{cccccc}
 0              & e            & -\nu       & d^c_{R}        & d^c_{G}          & d^c_{B}   \\
 -e        & 0            & e^c        & -u_{R}         & -u_{G}           & -u_{B}    \\
 \nu       & -e^{c}   & 0          & -d_{R}         & -d_{G}           & -d_{B}    \\
 -d^{c}_R   & u_R     & d_R   & 0              & u^c_{B}          & -u^c_{G}  \\
 -d^{c}_G   & u_G     & d_G   & -u^{c}_{B} & 0                & u^c_{R}   \\
 -d^{c}_B   & u_B     & d_B   & u^{c}_{G}  & -u^{c}_{R}   & 0         \\
\end{array} \right)$$

\noindent Here every component is a $5\times 5$ matrix which includes both left and right-charge-conjugated fields, plus, as we have seen, the gauge fields of $SL(2,\mathbb{C})$ and $SL(3,\mathbb{C})$ (the latters identical in every component). $G, R, B$ denote $SU(3)$ charges \emph{Green, Red, Blue}; $u$ denotes quark \emph{up, charm, top} depending on the family; $d$ denotes quark \emph{down, strange, bottom} depending on the family, $e$ denotes \emph{electron, muon or lepton }$\t$ depending on family, $\nu$ denotes neutrino and $^c$ indicates charge conjugation. It's straightforward to verify the correctness of their transformation laws under $SU(3)\otimes SU(2) \otimes U(1)$, respecting also the chirality of $SU(2)$.

\section{Second Quantization}

Let's promote $M$ to a quantum field operator and make the following decomposition in terms of creation/annihilation operators:
\ba M_{ij} &=& \fr 12\left(a_{ij} + b^\dag_{ij}\right) \qquad N_{ij} = a^\dag_{ij}a_{ij}\nonumber \\
    M^\dag_{ij} &=& \fr 12\left(a^\dag_{ij} +b_{ij}\right) \qquad N_{ji} = b^\dag_{ij}b_{ij}\nonumber \ea

\noindent The action of such operators over states is easily illustrated in the figures

\begin{figure}[h!]
\centering\includegraphics[width=0.7\textwidth ]{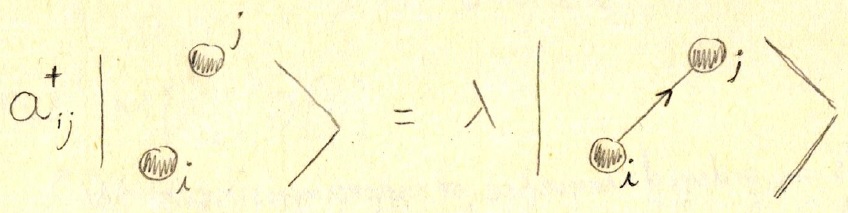}
\label{grafo}
\end{figure}
\begin{figure}[h!]
\centering\includegraphics[width=0.7\textwidth ]{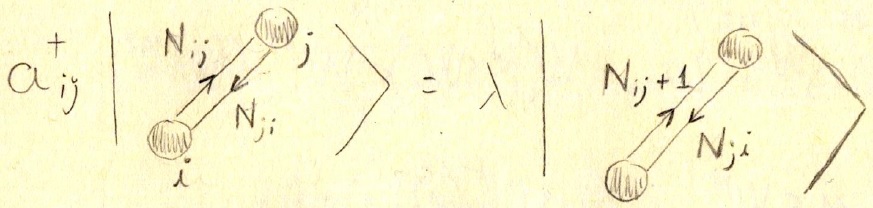}
\label{grafo}
\end{figure}
\begin{figure}[h!]
\centering\includegraphics[width=0.7\textwidth ]{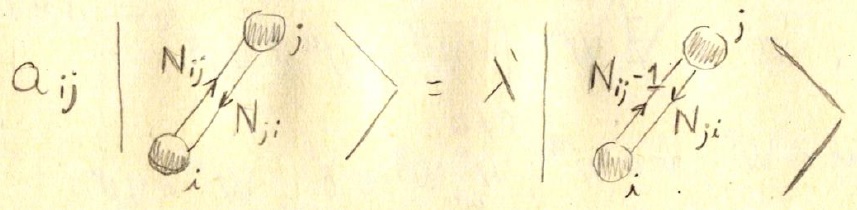}
\label{grafo}
\end{figure}
\begin{figure}[h!]
\centering\includegraphics[width=0.7\textwidth ]{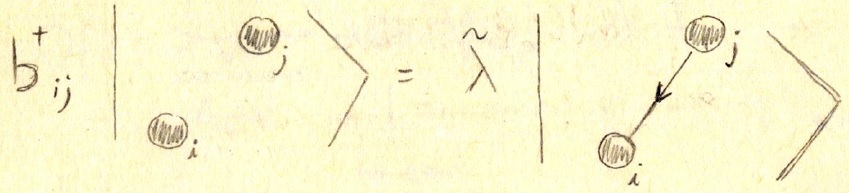}
\label{grafo}
\end{figure}
\begin{figure}[h!]
\centering\includegraphics[width=0.7\textwidth ]{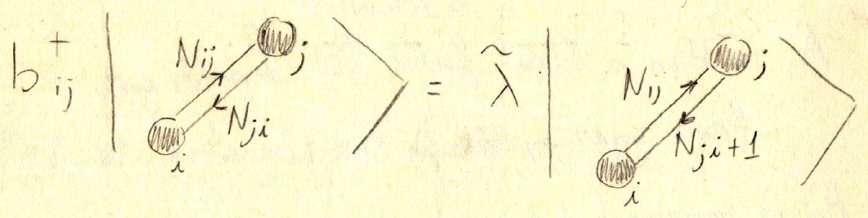}
\label{grafo}
\end{figure}
\begin{figure}[h!]
\centering\includegraphics[width=0.7\textwidth ]{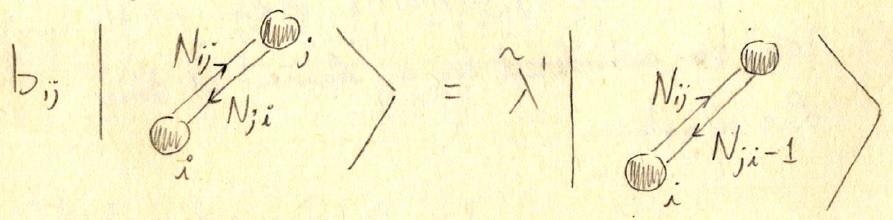}
\label{grafo}
\end{figure}
\newpage
\noindent Reasoning by analogy with loop gravity, we define an area operator as
$$S_{ij} = \{M_{ij},M^\dag_{ij}\} = \fr 12 \left(N_{ij}+N_{ji}\right)+\fr 12$$

\noindent We see that any two atoms are connected by a surface with area different from zero, independently from their \lq\lq classical'' distance. The minimal value for these areas is $\fr 12$ in natural unities, i.e. one half of Planck area. Can we use this small area to transmit information? We don't know.

An area $S_{ij}$ can be obtained in $2S_{ij}+1$ ways. For example:

\begin{figure}[h!]
\centering\includegraphics[width=1\textwidth ]{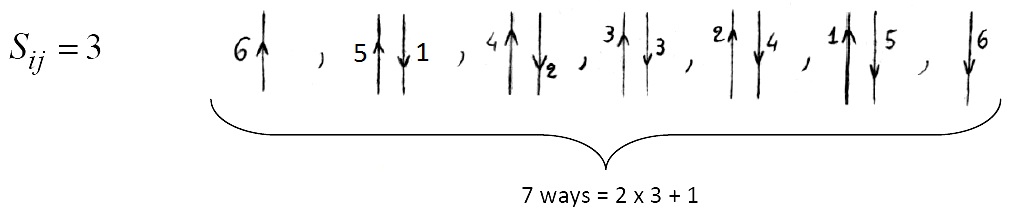}
\label{grafo}
\end{figure}

\noindent Hence $2S_{ij}+1$ is the \textbf{weight} of area $S_{ij}$.

\section{Path Integral}

A gauge invariant Path Integral can be defined as follows
\ba Z &=& \int DM \!\!\!\!\sum_{{possibly\pt choices} \atop {of\pt indep.\chi,\theta}} \!\!\! exp\left(\fr 1\s\sum_{indep.\chi,\chi'}\sum_{i,j} \na_{ij}[\chi]\na^*_{ji}[\chi']+\ldots\right.\nonumber \\
&& \left.\ldots +g_N \!\!\!\!\!\!\sum_{{indep.}\atop {\theta_1,\theta_2,\ldots,\theta_N}}\sum_{i,j,k,l,s,\ldots,u,w} \!\!\! \na_{ij}[\theta_1]\na^*_{jk}[\theta_2]\na_{kl}[\theta_3]\na^*_{ls}[\theta_4]\ldots \na_{uw}[\theta_{N-1}]\na^*_{wi}[\theta_N]\right)\nonumber \ea

\noindent Due to the limited space we have omitted to include $\phi$. For taking it into account it is sufficient to substitute $\sum_{indep.\chi}\na_{ij}[\chi]$ with $\phi_{ij} + \sum_{indep.\chi}\na_{ij}[\chi]$. The action dependence from $M$ is implicit inside $\na,\na^\dag$. We see that congruences $\chi,\theta$ take the place of coordinates, while matrix $\na$ behaves like an unified field. $g_N$ is simply a coupling. We can stop at $N=4$ to obtain all standard model terms.

\section{Explicit calculation of space-time dimensions}
\label{dimensions}

We consider a space (space-time) which contains all over $N$ atoms. We can easily define the number $n$ of independent non-diagonal connections:
$$ n = \fr{N(N-1)}{2}\langle P \rangle $$

\noindent with $\langle P \rangle $ the average probability for non-diagonal connections (where average is intended over all connections in a fixed state):
$$\langle P \rangle = \fr 2 {N(N-1)} \sum_{ij}|M'_{ij} M'_{ji}|^2 $$

\noindent $M'$ is equal to $M$ with diagonal elements taken as zero. Here we have considered \lq\lq classical'' connections, i.e. connections where both $i$ is connected to $j$ and $j$ is connected to $i$. The probability amplitude for such connections is obviously the product between probability amplitudes for $i$ to be connected to $j$ and for $j$ to be connected to $i$. Hence
$$n= \fr {N(N-1)}2 \fr 2 {N(N-1)} \sum_{ij}|M'_{ij} M'_{ji}|^2 = \sum_{ij}|M'_{ij} M'_{ji}|^2$$

\noindent Approximate the universe as a cubic lattice with step $L_p$. In presence of $d$ dimensions it must be true
$$ n = d\left( \fr L {L_P}\right)^d$$

\noindent where $L$ is the diameter of universe and $L_P$ is the fundamental length (the Planck length). Accordingly
$$d \left( \fr L {L_P}\right)^d = \sum_{ij}|M'_{ij} M'_{ji}|^2$$
$$\left\langle \sum_{ij}|M'_{ij} M'_{ji}|^2\right\rangle = \fr{\int dM e^{\fr 1 \sigma Tr\,M^2 +g Tr\,M^4}\sum_{ij}|M'_{ij} M'_{ji}|^2}{\int dM e^{\fr 1 \sigma Tr\,M^2 +g Tr\,M^4}}\quad\overset{g\sim 0}{\longrightarrow}\quad \fr{N(N-1)}{2}\left(\fr \sigma 2\right)^4$$

\noindent Here the average is intended over all the states. $\fr{N(N-1)}{2}$ is the number of terms which add up inside $\sum_{ij}|M'_{ij} M'_{ji}|^2$. The exponent $4$ is due to the fact that every $M_{ij}$ is an hypercomplex variable generated by $50$ units, of which $26$ are associated to bosonic degrees which contribute with $\left(\sigma/2\right)^{2\cdot 26}$, while the other $24$ are associated to fermionic degrees which - as well known - they invert the variance of distribution, contributing with $\left(\fr{2}{\sigma}\right)^{2\cdot 24}$. Please note that squared variance is here $\sigma/2$. Putting pieces together:
$$d\left(\fr{L}{L_P}\right)^d = \fr{N(N-1)}{32}\sigma^4 \cong \fr {N^2}{32} \sigma^4 = \left(\fr{L}{L_P}\right)^{2d} \fr {\sigma^4}{32}$$

\noindent Taking logarithm:
$$log\,d + d\,log\,\left(\fr L{L_P}\right) = 2d\,log\,\left(\fr L{L_P}\right) - log\,32 + 4\,log\,\sigma$$
$$d\,log\,\left(\fr L{L_P}\right) = -4\,log\,\sigma + log\,(32 d)$$

\noindent Finally
$$d = -\fr{4\,log\,\sigma}{log\,\left(\fr L{L_P}\right)}+\fr{log\,(32 d)}{log\,\left(\fr L{L_P}\right)}$$

\noindent Consider
$$[\sigma] = [mass^2] \Rightarrow \sigma \overset{!}{=} \sqrt \Lambda = \sqrt {10^{-122}} = 10^{-61}$$
This means that the cosmological constant $\Lambda$ defines the oscillation amplitude of universe around the classical configuration. Moreover
\ba && L \cong 14 \div 100 \cdot 10^9 l.y. \cong 1,4 \div 10 \cdot 10^{26} m \nonumber\\
&& L_P \cong 1,6 \cdot 10^{-35} m \qquad (4\cdot 10^{-35} \pt\text{by using}\pt L_P = \sqrt{2\pi G} \pt\text{instead of}\pt \sqrt G \nonumber \ea

\noindent The smallest value of $L$ is evaluated in the time direction where $L = c\cdot t_U = c\cdot 14\cdot 10^9 y = 14\cdot 10^9 l.y.$. The equation is solved for $d=4$. In fact
$$ 4 = \fr{4\cdot 61 + log\,(128)}{61\div 62} = \fr{246,11}{61\div 62} = 3,97\div 4,03$$

\noindent Note that this is the first calculation in all literature that considers $d$ as a free computable variable. Even more important: the result of computation is $4$, exactly the number of perceived dimensions.

You see that $d$ is roughly equal to twice $b - f$, so that the choice of algebra $\mathfrak{H}$ determines univocally the dimensions $d$ of space time. In general, for $\mathfrak{H} = gl(r, \mathbb{C})$ we have\footnote{We consider that in $d$ dimensions a Weyl spinor has $2^{d/2-1}$ complex components.}
$$f = 2^{d/2-1}*(r-2^{d/2-1})*2*2 = 2^{d/2+1}r - 2^d$$
$$b = 2*\left(2^{d/2-1}\right)^2 + 2*(r-2^{d/2-1})^2 = 2^{d} + 2 r^2 -2^{d/2+1}r$$
$$d = 2(b - f) = 2(2^{d+1}-2^{d/2+2}r +2r^2)$$
And so
$$d = 2^{d+2}-2^{d/2+3}r+4r^2$$
We can solve for $r$

$$r_{1,2} = 2^{d/2} \pm \fr{\sqrt{d}}2$$
The first solutions are then $(3,5)$ for $d=4$ and $(254,258)$ for $d=16$. You see that $r = 3$ and $r=5$ are the unique tractable solutions, one with one family and one with three families ($fam. = r-2^{d/2-1} = 2^{d/2-1} \pm \fr{\sqrt{d}}2$). This in turn suggests the non-existence of extra dimensions in the classical configuration.

\section{A comment on the vielbein field $e^\theta_i$}

If we consider $M$ as the fundamental field instead of $\na[\theta]$ (as we have done in the calculation of dimensionality), then vielbein field disappears. Despite the illusory character of such field, a quantum character for gravitational force is guaranteed by spin connections $A^i_{\theta i} = tr\,A_\theta$ inside $M$.

In this framework $e^\theta_i$ behaves just like a coupling and the Einstein equations have to result as renormalization equations, i.e. $\b$-functions for $e^\theta_i$ which depend from a tensorial energy scale $T_{\mu\nu}$. Thus the Hilbert-Einstein action would be an effective action and not a fundamental one. An easy way - although non rigorous - to deduce such behaviour, makes use of the cited duality with \emph{String Theory} and it will be exposed in a next release of this work.

\section{Conclusion}

\emph{AFT} represents a new approach to high energy physics which merges the quantum nature of matter/forces with the dynamical character of space-time suggested by general relativity. The bonus of such union is the automatic dynamization of the number of physical dimensions and fields families, whose most probable values are predicted by theory itself. \emph{AFT} includes also a non local part which describes the global structure of universe, its transitions and the entanglement phenomena. It posses a wide symmetry (then broken) which mixes fermionic and bosonic fields, like supersymmetry but with a much smaller number of never observed extrafields. Finally bosonic and fermionic degrees are not exactly balanced, leaving space for a small value cosmological constant.

The listed features are enough to stimulate new research and testing. Top it off, it has a clear low energy limit under which we can recover the Standard Model, so that it is perfectly compatible with whole the already verified high energy physics.

\end{document}